\documentclass{ws-procs961x669}            % default, citations in superscript
\begin{document}
\title{$U(1)$ local strings in generalized hybrid metric-Palatini gravity}

\author{Hilberto M. R. da Silva}

\address{Instituto de Astrofísica e Ciências do Espaço, Universidade do Porto, CAUP, Rua das Estrelas,
PT4150-762 Porto, Portugal and Centro de Astrofísica da Universidade do Porto,
Rua das Estrelas, PT4150-762 Porto, Portugal}

\author{Tiberiu Harko}

\address{Astronomical Observatory, 19 Ciresilor Street, 400487 Cluj-Napoca, Romania \\
Faculty of Physics, Babes-Bolyai University, 1 Kogalniceanu Street, 400084 Cluj-Napoca, Romania \\
School of Physics, Sun Yat-Sen University, Xingang Road,
510275 Guangzhou, People’s Republic of China}

\author{Francisco S. N. Lobo}
\address{Instituto de Astrofísica e Ciências do Espaço, Faculdade de Ciências da Universidade de Lisboa,
Edifício C8, Campo Grande, P-1749-016 Lisbon, Portugal}

\author{Jo\~{a}o Lu\'{i}s Rosa\!\!}

\address{Institute of Physics, University of Tartu, W. Ostwaldi 1, 50411 Tartu, Estonia}

\begin{abstract}
In this work we will explore $U(1)$ local cosmic string solutions in the context of the generalized hybrid metric-Palatini theory of gravity in its scalar-tensor representation. Using a general static cillindrically symmetric metric to find the dynamical equations for this particular case, we will simplify the equations by imposing boost invariance along $t$ and $z$ directions. The strings properties are determined by both scalar fields and by the effective potential, function of the scalar fields. While for some forms of the potential, the dynamical equations can be solved exactly, for more general forms of the potential the solutions are found numerically. Several stable string configurations were found, whose basic parameters depend essentially on the effective field potential, and on the boundary conditions. 
\end{abstract}


\bodymatter

\section{Introduction}\label{intro}

The primary motivation behind the investigation of hybrid metric-Palatini theories is the fact that these theories are able to overcome flaws of both the metric and the Palatini approaches to $f\left(R\right)$ gravity. Considering $f\left(R\right)$ in both these  formalism, one is able to model the late-time cosmic acceleration without the need for dark energy sources \cite{Sotiriou:2008rp}, but both approaches present profound drawbacks: the metric $f\left(R\right)$ was shown to be inconsistent with solar-system constraints unless chameleon mechanisms are considered \cite{Khoury:2003aq,Khoury:2003rn}, whereas the Palatini $f\left(R\right)$ gravity induces microscopic instabilities, surface singularities in polytropic star models, and is unable to describe the evolution of cosmological perturbations \cite{Olmo:2011uz, Gomez:2020rnq}. The HMPG, on the other hand, is capable of successfully unify both late-time cosmic acceleration period with the weak-field solar system dynamics free from chameleon mechanisms \cite{Harko:2011nh}, thus being a viable and relevant modification to GR. We refer the reader to Refs. \cite{Harko:2018ayt,Capozziello:2015lza,Harko:2020ibn} for recent reviews on the topic.

Inspired by the success of electroweak theory \cite{Weinberg:1967tq, Salam:1959zz,Salam:1964ry}, which unifies the weak and electromagnetic interaction under the gauge group $SU(2) \times U(1)$ at a scale of around $10^2 GeV$, Grand Unified Theories (GUT) propose the unification of electroweak and strong interactions under a more general symmetry that takes place at higher energy scales, around $10^{16} GeV$, the Grand Unification Scale. GUT theories are supported by the observation that the coupling ``constants" of the Standard Model for Particle Physics seem to slowly vary with the energy scale, converging to a common value at the Grand Unification Scale \cite{Amaldi:1991zx}.

These symmetries presented at higher energies are spontaneously broken as the system lowers its energy state. In several GUT scenarios proposed, a universal covering group $G$, which would be effective above the GUT scale, would spontaneously break into the Standard Model $SU(3) \times SU(2) \times U(1)$, where $SU(3)$ is the symmetry group of quantum chromodynamics, describing the strong interaction, and $SU(2) \times U(1)$ is the aforementioned electroweak group. 

These phase transitions may have left behind some relics that can help shed some light into earlier times of our Universe \cite{Jeannerot:2003qv}.
These relics are known as topological defects and are a well known, and studied, phenomena in physics, particularly in the context of condensed matter (namely metal crystallization \cite{Mermin:1979zz}, liquid crystals \cite{Chuang:1991zz}, superfluid helium-3 and helium-4 \cite{Salomaa:1987zz}, and superconductivity \cite{Abrikosov:1956sx}).

The underlying idea behind topological defect formation is the one of Spontaneous symmetry breaking, which is the principle behind the Higgs-Englert mechanism \cite{Higgs:1964pj}.

Cosmic strings are one of the possible topological defects formed after spontaneous symmetry breaking (SSB) during phase transitions in the history of the Universe.

The type of strings to be considered in this work are local $U(1)$ cosmic strings, which are an extension of the global $U(1)$ strings to include gauge fields. Local strings differ from the global cosmic strings in what concerns the symmetry that is effective above the spontaneous breaking scale; in the case of local strings, the lagrangian remains invariant under local transformations of the type $\phi(x) \longrightarrow e^{i\alpha(x)}\phi (x)$.

The study of the properties and dynamics of cosmic strings in the context of modified theories of gravity is crucial in the advent of powerful observatories, such as LISA, as it may allow us to constrain both Modified Gravity theories and Grand Unified theories.

\section{Generalized Hybrid metric-Palatini Gravity}
The Hybrid metric-Palatini gravity theory in its generalized version ca be cast with the following action:
\begin{equation}\label{genac}
S=\frac{1}{2\kappa^2}\int_\Omega\sqrt{-g}f\left(R,\cal{R}\right)d^4x+
\int_\Omega\sqrt{-g}\;{\cal L}_m d^4x,
\end{equation}
where $\kappa^2 \equiv 8\pi G/c^4$, $G$ is the gravitational constant and $c$ the speed of light, $\Omega$ is the spacetime manifold described by a system of coordinates $x^a$, $g$ is the determinant of the spacetime metric $g_{ab}$, where Latin indices run from 0 to 3, $R=g^{ab}R_{ab}$ is the Ricci scalar of the metric $g_{ab}$ and where $R_{ab}$ is the Ricci tensor, $\mathcal{R}\equiv\mathcal{R}^{ab}g_{ab}$ is the Palatini Ricci scalar, obtained from the Palatini-Ricci tensor $\mathcal R_{ab}$ that is constructed from an independent connection $\hat\Gamma^c_{ab}$ in the usual form as $\mathcal{R}_{ab}=\partial_c\hat\Gamma^c_{ab}-\partial_b\hat\Gamma^c_{ac}+\hat\Gamma^c_{cd}\hat\Gamma^d_{ab}-\hat\Gamma^c_{ad}\hat\Gamma^d_{cb}$, where $\partial_a$ denotes partial derivatives with respect to the coordinates $x^a$,
$f\left(R,\cal{R}\right)$ is a well-behaved function of $R$ and $\cal{R}$, and ${\cal L}_m$ is the matter Lagrangian density, that is taken to be minimally coupled to the metric $g_{ab}$. Equation \eqref{genac} depends on the metric $g_{ab}$ and the independent connection $\hat\Gamma^c_{ab}$, and thus two equations of motion can be obtained.

Varying Eq. \eqref{genac} with respect to the metric $g_{ab}$ we obtain the modified field equations
\begin{eqnarray}
\frac{\partial f}{\partial R}R_{ab}+\frac{\partial f}{\partial \mathcal{R}}\mathcal{R}_{ab}-\frac{1}{2}g_{ab}f\left(R,\cal{R}\right)
 %  \nonumber \\
-\left(\nabla_a\nabla_b-g_{ab}\Box\right)\frac{\partial f}{\partial R}=\kappa^2 T_{ab},\label{field1}
\end{eqnarray}
where $\nabla_a$ denotes covariant derivatives and $\Box=\nabla^a\nabla_a$ the d'Alembert operator, both with respect to $g_{ab}$, and $T_{ab}$ is the energy-momentum tensor defined as usual:
\begin{equation}
T_{ab}=-\frac{2}{\sqrt{-g}}\frac{\delta(\sqrt{-g}\,{\cal L}_m)}{\delta(g^{ab})} ~.
 \label{defSET}
\end{equation}
By varying Eq. \eqref{genac} with respect to the independent connection $\hat\Gamma^c_{ab}$ we obtain the equation
\begin{equation}
\hat\nabla_c\left(\sqrt{-g}\frac{\partial f}{\partial \cal{R}}g^{ab}\right)=0 \,,
\label{eqvar1}
\end{equation}
where $\hat\nabla_a$ is the covariant derivative written in terms of the independent connection $\hat\Gamma^c_{ab}$. Since $\sqrt{-g}$ is a scalar density of weight 1, then $\hat\nabla_c \sqrt{-g}=0$ and Eq. \eqref{eqvar1} can be rewritten in the form $\hat\nabla_c\left(\frac{\partial f}{\partial \cal{R}}g^{ab}\right)=0$. So they is a new metric, conformally related to the metric $g_{ab}$, $h_{ab}$ through
\begin{equation}
h_{ab}=g_{ab} \frac{\partial f}{\partial \cal{R}} \,,
\label{hab}
\end{equation}
the independent connection is Levi-Civita of the metric $h_{ab}$, i.e., $\hat\Gamma^c_{ab}$ can be written as
\begin{equation}
\hat\Gamma^a_{bc}=\frac{1}{2}h^{ad}\left(\partial_b h_{dc}+\partial_c h_{bd}-\partial_d h_{bc}\right)\,.
\end{equation}

%%%%%%%%%%%%%%%%%%%%%%%%%%%%%%%%%%%%%%%%%%%%%%%%%%%%%%%%%%%%%
\subsection{Scalar-tensor representation of generalized HMPG with matter}\label{str}
%%%%%%%%%%%%%%%%%%%%%%%%%%%%%%%%%%%%%%%%%%%%%%%%%%%%%%%%%%%%%

The Generalized version of HMP gravity can be recast in a dinamically equivalent scalar-tensor representation. In this case, the extra scalar degrees of freedmon of the theory are explicitly represented by a pair of scalar fields. To achieve this representation of the theory, we introduce two auxiliary fields $\alpha$ and $\beta$ into Eq. \eqref{genac} and rewrite it in the form
\begin{eqnarray}
S=\frac{1}{2\kappa^2}\int_\Omega \sqrt{-g}\left[f\left(\alpha,\beta\right)+\frac{\partial f}{\partial \alpha}\left(R-\alpha\right)
   % \nonumber  \\
+\frac{\partial f}{\partial\beta}\left(\cal{R}-\beta\right)\right]d^4x+\int_\Omega\sqrt{-g}\;{\cal L}_m d^4x.
\label{gensca}
\end{eqnarray}
If $\alpha=R$ and $\beta=\mathcal R$ one recovers Eq. \eqref{genac}. Defining two scalar fields as $\varphi=\partial f(\alpha,\beta)/\partial\alpha$ and $\psi=-\partial f(\alpha,\beta)/\partial\beta$ (negative sign in $\psi$ is used to avoid the presence of ghosts), Eq. \eqref{gensca} takes the form
\begin{eqnarray}
S=\frac{1}{2\kappa^2}\int_\Omega \sqrt{-g}\left[\varphi R-\psi\mathcal{R}-V\left(\varphi,\psi\right)\right]d^4x
%	\nonumber \\
+\int_\Omega\sqrt{-g}\;{\cal L}_m d^4x,
  \label{action3}
\end{eqnarray}
where the function $V\left(\varphi,\psi\right)$ plays the role of an interaction potential between both scalar fields and it is defined as
\begin{equation}\label{potential}
V\left(\varphi,\psi\right)=-f\left(\alpha,\beta\right)+\varphi\alpha-\psi\beta \,.
\end{equation}
Recalling that $h_{ab}$ and $g_{ab}$ in Eq. \eqref{hab} are conformally related, we can now write $h_{ab}=-\psi\, g_{ab}$ by taking into consideration the definition of $\psi$. We can derive a relationship between $R$ and $\mathcal R$ as
\begin{equation}\label{confrt}
\mathcal{R}=R+\frac{3}{\psi^2}\partial^a \psi\partial_a \psi-
\frac{3}{\psi}\Box\psi\,,
\end{equation}
which can be used to remove $\mathcal R$ from Eq. \eqref{action3} and gives the final form of the action
\begin{eqnarray}\label{genacts2}
S=\frac{1}{2\kappa^2}\int_\Omega \sqrt{-g}\Big[\left(\varphi-\psi\right) R
-\frac{3}{2\psi}\partial^a\psi\partial_a\psi
 %    \nonumber  \\
 -V\left(\varphi,\psi\right)\big]d^4x+\int_\Omega\sqrt{-g}\;{\cal L}_m d^4x.
\end{eqnarray}

Equation \eqref{genacts2} is now dependent on three variables, namely, the metric $g_{ab}$ and both scalar fields $\varphi$ and $\psi$. Varying Eq. \eqref{genacts2} with respect to the metric $g_{ab}$ yields the modified field equations in the scalar-tensor representation. Varying the action \eqref{genacts2} with respect to the metric $g_{ab}$
provides the following gravitational equation
\begin{eqnarray}
\left(\varphi-\psi\right) G_{ab}=\kappa^2T_{ab}
+\nabla_a\nabla_b
\varphi-\nabla_a\nabla_b\psi
+\frac{3}{2\psi}\partial_a\psi\partial_b\psi
	\nonumber  \\
 -\left(\Box\varphi-\Box\psi+\frac{1}{2}V+\frac{3}{4\psi}
\partial^c\psi\partial_c\psi\right)g_{ab}\,.
\label{genein2}
\end{eqnarray}

The Klein-Gordon Equations for the scalar fields $\varphi$ and $\psi$ can be obtained by variations of Eq. \eqref{genacts2} with respect to these fields, which results in
\begin{equation}\label{genkgi}
\Box\varphi+\frac{1}{3}\left(2V-\psi V_\psi-\varphi V_\varphi\right)
=\frac{\kappa^2T}{3}\,,
\end{equation}
\begin{equation}\label{genkg}
\Box\psi-\frac{1}{2\psi}\partial^a\psi\partial_a\psi-\frac{\psi}{3}
\left(V_\varphi+V_\psi\right)=0\,,
\end{equation}
respectively.

From Eq.\eqref{genacts2}, the coupling between the scalar fields and the Ricci scalar is the combination $\varphi-\psi$. We thus introduce a redefinition of the scalar field $\varphi$ as $\xi^2=\varphi-\psi$. With this redefinition, any solution for which $\xi$ is a real function preserves the positivity of the coupling $\left(\varphi-\psi\right)R$. Equations \eqref{genein2} to \eqref{genkg} thus become
\begin{equation}\label{genein3}
\xi^2G_{ab}=\kappa^2 T_{ab}+\nabla_a\nabla_b\xi^2+\frac{3}{2\psi}\partial_a\psi\partial_b\psi-\left(\Box\xi^2+\frac{1}{2}\bar V+\frac{3}{4\psi}\partial^c\psi\partial_c\psi\right)g_{ab},
\end{equation}
\begin{equation}\label{genkgxi}
\Box\xi^2+\frac{1}{2\psi}\partial^a\psi\partial_a\psi+\frac{1}{6}\left(4\bar V-\xi \bar V_\xi\right)=\frac{\kappa^2T}{3},
\end{equation}
\begin{equation}\label{genkgpsi}
\Box\psi-\frac{1}{2\psi}\partial^a\psi\partial_a\psi-\frac{\psi}{3}\left(\frac{1}{2\xi}\bar V_\xi+\bar V_\psi\right)=0,
\end{equation}
where $\bar V\left(\xi,\psi\right)$ is the potential in terms of $\xi$ and $\psi$.  We will now use the set of equations (\ref{genein3})--(\ref{genkgpsi}) to find cosmic string solutions. Finally, one can also obtain a relationship between the potential $\bar V$ and the function $f\left(R,\mathcal R\right)$ from Eq.\eqref{potential} as
\begin{equation}\label{Vin}
\bar V\left(\xi,\psi\right)=-f\left(R,\mathcal R\right)+\xi^2 R+\psi\left(R-\mathcal R\right),
\end{equation}
This equation becomes a PDE for $f\left(R,\mathcal R\right)$ by replacing $\psi=f_\mathcal R$ and $\xi^2=f_R+f_\mathcal R$.

Hence Eq.~(\ref{Vin}) becomes
\begin{equation}\label{Vin1}
V\left( \sqrt{\frac{\partial f\left( R,\mathcal{R}\right) }{\partial R}+\frac{\partial
f\left( R,\mathcal{R}\right) }{\partial \mathcal{R}}},\frac{\partial f\left( R,\mathcal{R}\right) }{%
\partial \mathcal{R}}\right) =-f\left( R,\mathcal{R}\right) +R \frac{\partial
f\left( R,\mathcal{R}\right) }{\partial R}+\mathcal R\frac{\partial f\left( R,\mathcal{R}\right) }{%
\partial \mathcal{R}}.
\end{equation}

\section{Dynamical equations of local strings in Hybrid metric-palatini using Vilenkin's approximation}
Using Vilenkin's approximation \cite{Vilenkin:1981zs}, the energy-momentum tensor of an infinite straight cosmic string can be put as
\begin{equation}\label{string}
    T^t_t=T^z_z=-\sigma(r) \,,
\end{equation}
where $\sigma$ is the string tension.
We further assume cylindrical symmetry with a general metric of the form:
\begin{equation}\label{metric}
    ds^2=-e^{2(K-U)}dt^2+e^{2(K-U)}dr^2+e^{-2U}W^2d\theta^2+e^{2U}dz^2,
\end{equation}
where $t$, $r$, $\theta$ and $z$ denote the time, radial, angular and axial cylindrical coordinates, respectively, and $K$, $U$ and $W$ are functions of $r$ alone.

It is possible to show that the energy conservation equation still holds (as matter fields couple only minimally with curvature), i.e.,
\begin{equation}\label{conservation}
    \nabla_a T^{a}{}_{b}=0
\end{equation}
which provides $K'\sigma=0$, this implies that $K'=0$, where the prime represents a differentiation w.r.t. $r$. Thus, we consider from now on that $e^K=1$.

Considering that this type of strings preserve boost invariance along $t$ and $z$ \cite{Vilenkin:1981zs}, this requires $U=0$. Hence the only surviving non-trivial metric tensor component is $g_{\theta \theta}=W^2(r)$, and so the metric of the cosmic string reduces to
\begin{equation}
\label{metrn}
ds^2=-dt^2+dr^2+W^2(r)d\theta^2+dz^2.
\end{equation}

With this simplifications, the gravitational field equations simplify considerably. Equation \eqref{genein3} provides three independent field equations, which are
\begin{eqnarray}\label{fieldtt}
\xi^2\frac{W''}{W}+2\xi\xi'\frac{W'}{W}
%	\nonumber \\
+\frac{3\psi'^2}{4\psi}+2\left(\xi'^2+\xi\xi''\right)+\frac{\bar V}{2}=-\kappa^2\sigma,
\end{eqnarray}
\begin{equation}\label{field11}
2\xi\xi'\frac{W'}{W}-\frac{3\psi'^2}{4\psi}+\frac{\bar V}{2}=0,
\end{equation}
\begin{equation}\label{field22}
2\left(\xi'^2+\xi\xi''\right)+\frac{3\psi'^2}{4\psi}+\frac{\bar V}{2}=\frac{d^2}{dr^2}\xi^2+\frac{3\psi'^2}{4\psi}+\frac{\bar V}{2}=0 \,,
\end{equation}
whereas the scalar field equations for $\xi$ and $\psi$, given by Eqs. \eqref{genkgxi} and \eqref{genkgpsi}, give
\begin{equation}\label{eomphi}
2\left(\xi'^2+\xi\xi''\right)+2\xi\xi'\frac{W'}{W}+\frac{\psi'^2}{2\psi}+\frac{1}{6}\left(\bar V-\xi \bar V_\xi\right)=-\frac{2\kappa^2}{3}\sigma,
\end{equation}
\begin{equation}\label{eompsi}
\psi''+\frac{W'}{W}\psi'-\frac{\psi'^2}{2\psi}-\frac{\psi}{3}\left(\bar V_\psi+\frac{1}{2\xi}\bar V_\xi\right)=0 \,.
\end{equation}

In the system of Eqs. \eqref{fieldtt}-\eqref{eompsi} only four are linearly independent. Given its complexity, we chose to discard Eq. \eqref{fieldtt} from the analysis, and proceed with the four linearly independent equations.

An equation for the potential $\bar V$ can be obtained by summing the field equations in Eqs. \eqref{field11} and \eqref{field22}, yielding
\begin{equation}\label{eq28}
\bar V=-2\left(\xi'^2+\xi\xi''\right)-2\xi\xi'\frac{W'}{W}.
\end{equation}
This equation is particularly useful to obtain an equation for $W'$ in terms of the scalar fields $\xi$ and $\psi$ and their derivatives after setting an explicit form of the potential $\bar V$.

The system of basic equations describing the structure of a cosmic string can thus be reformulated in the form of a first-order dynamical system. By defining $\alpha =\xi ^2$, and  by introducing two extra dynamical variables as $u=\alpha '$ and $v=\psi'$, the dynamical system takes the form
\begin{equation}\label{eq1}
\frac{d\alpha }{dr}=u, \qquad \frac{d\psi }{dr}=v,
\end{equation}%
\begin{equation}\label{eq2}
\frac{dW}{dr}=\frac{1}{u}\left( \frac{3v^{2}}{4\psi }-\frac{\bar{V}}{2}\right) W,
\end{equation}%
\begin{equation}\label{eq3}
\frac{du}{dr}=-\frac{3v^2}{4\psi}-\frac{\bar{V}}{2},
\end{equation}%
\begin{equation}\label{eq4}
\frac{dv}{dr}=-\frac{v}{u}\left(\frac{3v^2}{4\psi}-\frac{\bar{V}}{2}\right)+\frac{v^2}{2\psi}+\frac{\psi }{3}\left( \bar{V}_{\psi }+\frac{1}{2\sqrt{\alpha}}\bar{V}_{\sqrt{\alpha} }\right) ,
\end{equation}
where Eq. \eqref{eq1} is the explicit definition of $u$ and $v$, and Eqs. \eqref{eq2}--\eqref{eq4} are reformulations of Eqs. \eqref{field11}, \eqref{eomphi}, and \eqref{eompsi}, respectively. Once the form of the potential $\bar{V}(\xi,\psi)$ is specified, the system of Eqs. (\ref{eq1})-(\ref{eq4}) represents a system of ordinary, strongly nonlinear, differential equations for the variables $\left(\alpha =\xi ^2,\psi, W, u,v\right)$. To solve this system, one has to impose a set of boundary conditions at some radius $r=r_0$, i.e., $\alpha \left(r_0\right)=\alpha _0$, $\psi \left(r_0\right)=\psi _0$, $W \left(r_0\right)=W _0$, $u \left(r_0\right)=u _0$, and $v \left(r_0\right)=v_0$, respectively, which specify the boundary values of the variables on, or nearby the string axis. Moreover, we will also impose the condition $u\left(r_0\right)\neq v\left(r_0\right)$. Once the system is solved, the string tension can be obtained from Eq.~(\ref{eomphi}), and it is given by
\begin{equation}
\frac{2}{3}\kappa ^2 \sigma=\bar{V}-\frac{v^2}{2\psi}-\frac{1}{6}\left(\bar{V}-\sqrt{\alpha}\bar{V}_{\sqrt{\alpha}}\right).
\end{equation}

An important physical characteristic of the string-like objects is their mass per unit length $m_s$, defined as
\begin{eqnarray}
m_s\left(R_s\right)=\int_0^{2\pi}{d\theta}\int_0^{R_s}{\sigma (r)W(r)dr}
	= 2\pi \int_0^{R_s}{\sigma (r)W(r)dr},
\end{eqnarray}
where $R_s$ is the radius of the string, defined as the distance from the center where the string tension vanishes, $\sigma \left(R_s\right)=0$, and $\sigma (r)\equiv 0, \forall r\geq R_s$. Note that, in general, the solutions obtained for $\sigma$ do not satisfy the property $\sigma (r)\equiv 0, \forall r\geq R_s$, and albeit being out of the scope of this article, this condition must be imposed manually by performing a matching between the string spacetime and an exterior cosmological spacetime. This matching must be performed via the use of the junction conditions of the theory, previously used in \cite{Rosa:2018jwp}. 

Using Eqs.~(\ref{field22}) and Eq.~(\ref{fieldtt}) the mass per unit length of the string can be expressed as
\begin{equation}
\kappa ^2 m_s\left(R_s\right)=3\pi\int_0^{R_s}{\left[\bar{V}-\frac{v^2}{2\psi}-\frac{1}{6}\left(\bar{V}-\sqrt{\alpha}\bar{V}_{\sqrt{\alpha}}\right)\right]Wdr}.
\end{equation}
\section{Solutions to the dynamical equations with specific potentials}

In this section we will apply the set of equations deduced on the previous section to different potential configurations, for a more complete set of possible potential configurations, we refer the reader to the original article Ref. \citenum{Rosa:2021zbk}.

\subsection{Constant potential}

First, we will consider the constant potential, where $V$ is a constant, so that $V=\Lambda={\rm constant}$.  In this case Eq.~(\ref{Vin1}) takes the form
\begin{equation}
-f\left( R,\mathcal{R}\right) +R \frac{\partial
f\left( R,\mathcal{R}\right) }{\partial R}+\mathcal R\frac{\partial f\left( R,\mathcal{R}\right) }{%
\partial \mathcal{R}}=\Lambda,
\end{equation}
and it has the general solution
\begin{equation}
f\left( R,\mathcal{R}\right)=R g\left(\frac{\mathcal R}{R}\right)+\mathcal R h\left(\frac{R}{\mathcal R}\right)-\Lambda,
\end{equation}
where $g$ and $h$ are arbitrary functions.

For a constant potential  Eq.~(\ref{eompsi}) simplifies to
\begin{equation}\label{eq51}
\frac{W'}{W}=-\frac{\psi''}{\psi '}+\frac{1}{2}\frac{\psi '}{\psi},
\end{equation}
which allows us to write Eq.~(\ref{field11}) in the form
\begin{eqnarray}
\hspace{-0.5cm}\frac{d}{dr}\xi ^2&=&\frac{3\psi '^2/4\psi-\Lambda/2}{W'/W}=\frac{3\psi '^2/4\psi-\Lambda/2}{-\psi ''/\psi'+(1/2)\psi'/\psi}.\label{eqVL}
\end{eqnarray}
To facilitate the analysis, we introduce now a new function $h=\psi '^2/\psi$. The radial derivative of this function can be written in terms of $\psi$ and its derivatives as
  \begin{equation}
  h'=2h\left(\frac{\psi ''}{\psi '}-\frac{1}{2}\frac{\psi '}{\psi}\right).
  \end{equation}

This definition allows us to rewrite Eq. \eqref{eqVL} in terms of $h$ as
  \begin{equation}\label{eq55}
 \frac{d}{dr}\xi ^2=-\frac{(3/4)h-\Lambda/2}{h'/2h},
  \end{equation}
which can then be differentiated with respect to $r$ and inserted into Eq.~(\ref{field22}) to cancel the dependency in $d^2\xi ^2/dr^2$ and $\psi''$. As a result, we obtain an equation depending solely in $h$ of the form
\begin{equation}\label{eq58}
\frac{(2 \Lambda -3 h) \left(3 h'^2-2 h h''\right)}{4 h'^2}=0,
\end{equation}
This equation is undefined for $h'=0$. Thus, we will ignore the solution corresponding to $h=2\Lambda/3={\rm constant}$, giving $h'=0$, and $W={\rm constant}$. The general solution of Eq.~(\ref{eq58}) is given by
\begin{equation}\label{solh}
h(r)=\frac{c_2}{\left(r+2
   c_1\right)^2},
\end{equation}
where $c_1$ and $c_2$ are arbitrary integration constants. Recalling that $h'=\psi'^2/\psi$, Eq.~\eqref{solh} becomes a separable ODE for $\psi$ which can be directly integrated and provides the general solution
\begin{equation}\label{solpsiVL}
\psi (r)=\left[c_3\pm \frac{\sqrt{c_2}}{2}\ln \left(r+2c_1\right)\right]^2,
\end{equation}
where $c_3$ is an arbitrary integration constant. Inserting Eq.~\eqref{solpsiVL} into Eq.~(\ref{eq55}) and integrating gives for $\xi ^2$ the expression
\begin{eqnarray}
\xi^2 (r)=\xi _0^2
	+\frac{1}{4} \left[3 c_2 \ln \left(r+2 c_1\right)-\Lambda  r \left(r+4 c_1\right)\right],
\end{eqnarray}
where $\xi _0^2$ is an integration constant. Inserting Eq. \eqref{solpsiVL} into Eq. \eqref{eq51} we obtain the solution for $W$
\begin{equation}
W(r)=W_0\left(r+2c_1\right),
\end{equation}
where $W_0$ is a constant of integration. Hence the cosmic string metric tensor component $W(r)$ is the same in both $V=0$ and $V=\Lambda$ cases. Finally, the string tension can be computed via Eq. \eqref{eomphi}, leading to
\begin{equation}
\kappa ^2\sigma =\frac{\Lambda }{2}-\frac{3 c_2}{4 \left(2 c_1+r\right)^2}.
\end{equation}

On the string axis $r=0$ we obtain for the string tension the value
\begin{equation}
\kappa ^2\sigma _0=\kappa ^2 \sigma (0)=\frac{\Lambda}{2}-\frac{3c_2}{16c_1^2}.
\end{equation}
If we enforce the positivity of the string tension, we must have $c_2/c_1^2<8\Lambda /3$ on the integration constants.

By a careful choice of the integration constants, and by assuming $\Lambda >0$, the string tension can be made positive in this model for all $r>0$. Moreover, $\lim_{r\rightarrow \infty}\sigma (r)=\Lambda/2\kappa ^2$, and hence at infinity the string tension becomes equal to the cosmological constant. However, in this case one can obtain a finite radius string configuration, with the radius $R_s$ determined by the condition $\sigma \left(R_s\right)=0$, and given by
\begin{equation}
R_s=\sqrt{\frac{3}{2}\frac{c_2}{\Lambda}}-2c_1.
\end{equation}
For a positive string tension at the origin $r=0$, the string radius is also positive.

As for the mass of the string we obtain
\begin{eqnarray}
m_s\left(R_s\right)=\frac{2\pi W_{0}}{\kappa ^{2}}\Bigg\{ \frac{1}{2}\Lambda \left[
R_{s}+c_{1}(\Lambda W_{0}+2)-\frac{3c_{2}W_{0}}{8c_{1}}\right]
\nonumber \\
	- \frac{%
6c_{1}c_{2}}{8c_{1}\left[ R_{s}+c_{1}(\Lambda W_{0}+2)\right] -3c_{2}W_{0}}%
\Bigg\}.
\end{eqnarray}
By an appropriate choice of the integration constants, giving the boundary conditions of the fields $\varphi$ and $\psi$ for $r=0$, one can always satisfy the condition $m_s\left(R_s\right)>0, \forall R_s$. In the limit $R_s\rightarrow \infty$, we obtain $m_s\left(R_s\right)\approx \left(\pi W_0\Lambda/\kappa  ^2\right)R_s$, that is, for large distances the mass of the string linearly increases with its radius.

\subsection{$\bar{V}\left(\xi,\psi\right)=\bar{V}_0\xi ^2\psi ^2$}

We will now investigate the dynamical equations considering a potential of the form $\bar{V}$ is given by $\bar{V}=\bar{V}_0\xi ^2\psi ^2=\bar{V_0}\alpha \psi ^2$, with $\bar{V}_0$ constant. Equation (\ref{Vin1}) then becomes
\begin{equation}
-f\left( R,\mathcal{R}\right) +R \frac{\partial
f\left( R,\mathcal{R}\right) }{\partial R}+\mathcal R\frac{\partial f\left( R,\mathcal{R}\right) }{%
\partial \mathcal{R}}=\bar V_0 \frac{\partial f\left( R,\mathcal{R}\right) }{%
\partial \mathcal{R}}\left(\frac{\partial f\left( R,\mathcal{R}\right) }{%
\partial \mathcal{R}}+\frac{\partial f\left( R,\mathcal{R}\right) }{%
\partial R}\right)
\end{equation}
and a particular solution for the function $f\left(R,\mathcal R\right)$ is
\begin{equation}
f\left(R,\mathcal R\right)=\sqrt{\frac{R}{\bar V_0}}\left(R-\mathcal R\right).
\end{equation}

For this potential the field equations describing the string-like structure take the form
\begin{equation}\label{eq1c}
\frac{d\alpha }{dr}=u, \qquad \frac{d\psi }{dr}=v,
\end{equation}%
\begin{equation}\label{eq2c}
\frac{dW}{dr}=\frac{1}{2u}\left( \frac{3v^{2}}{2\psi }-\bar{V}_0\alpha \psi ^2\right) W,
\end{equation}%
\begin{equation}\label{eq3c}
\frac{du}{dr}=-\frac{3v^2}{4\psi}-\frac{\bar{V}_0\alpha \psi^2}{2},
\end{equation}%
\begin{equation}\label{eq4c}
\frac{dv}{dr}=-\frac{v}{u}\left(\frac{3v^2}{4\psi}-\frac{\bar{V}_0\alpha}{2}\right)+\frac{v^2}{2\psi}+\frac{2V_0 }{3} \psi^2\left(\alpha +\frac{\psi}{2}\right).
\end{equation}

For this model the string tension is given by
\begin{equation}
\frac{2\kappa ^2}{3}\sigma=\frac{7}{6}\bar{V}_0\alpha \psi^2-\frac{v^2}{2\psi}.
\end{equation}

\begin{figure*}[htbp]
\centering
\includegraphics[width=8.3cm]{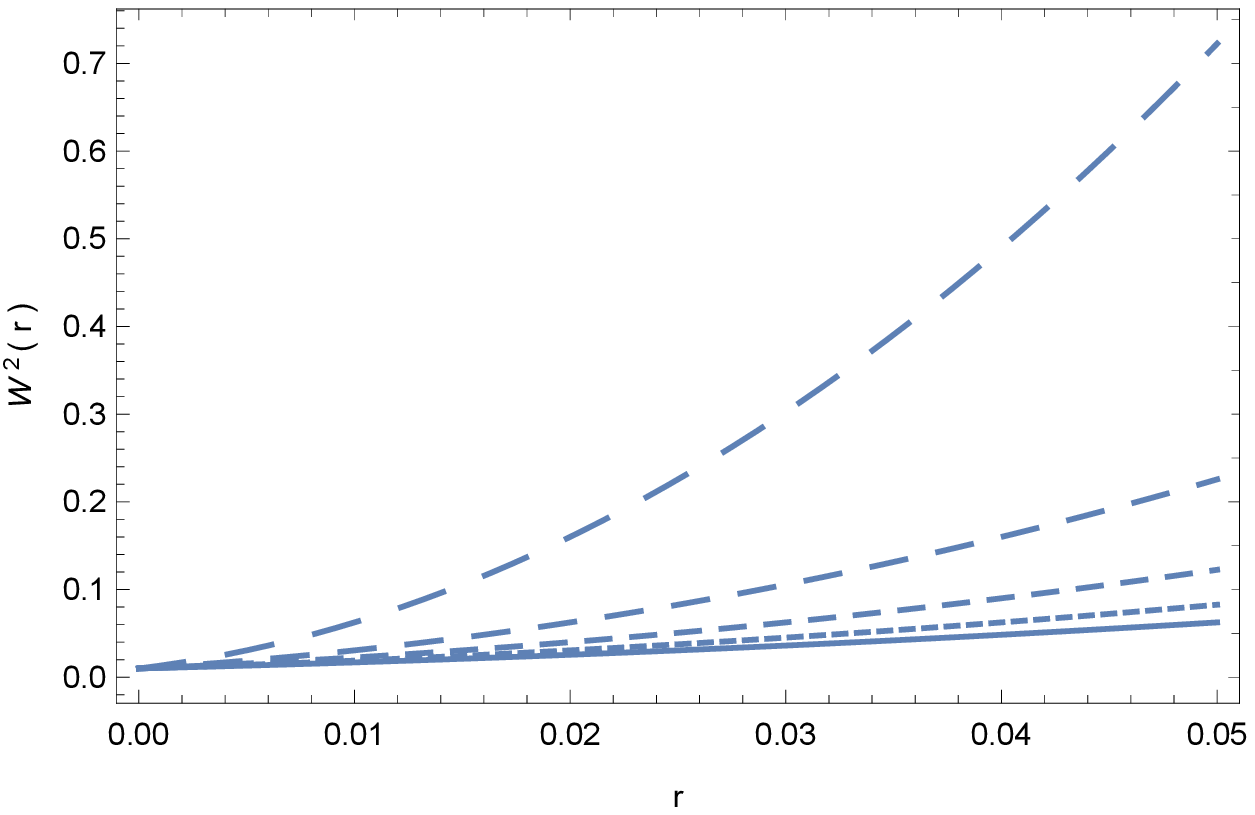}
\includegraphics[width=8.3cm]{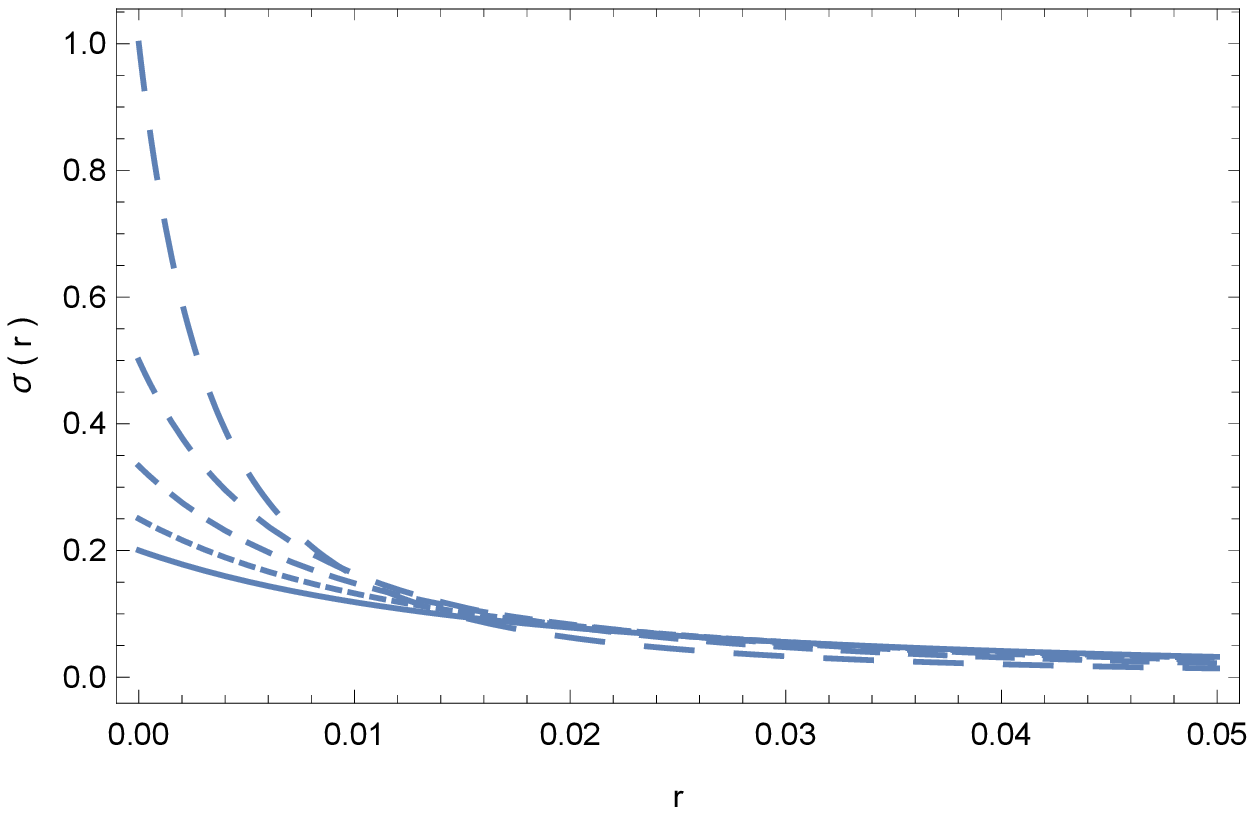}
\caption{Variations of the metric function  $W^2(r)$ (left panel), and of the string tension $\sigma (r) $ (right panel) as a function of $r$ (with all quantities in arbitrary units) for the $\bar{V}(\xi,\psi)=\bar{V}_0\xi ^2\psi^2$ potential, for $\psi _0=-0.025$ (solid curve), $\psi _0=-0.020$ (dotted curve),
 $\psi _0=-0.015$ (short dashed curve), $\psi _0=-0.01$ (dashed curve), and $\psi _0=-0.005$ (long dashed curve), respectively. For $\bar{V}_0$ we have adopted the value $\bar{V}_0=10$, while the boundary conditions used to numerically integrate the field equations are $u_0=-0.01$, $\alpha_0=0.025$, $W(0)=0.10$, and $v_0=0.10$, respectively. }
\label{fig6}
\end{figure*}

We can see both the metric function $W^2$ and the string tension $\sigma$ in Fig.~\ref{fig6}, for a varying initial condition $\psi '(0)=\psi_0$, while fixing all others initial conditions. In this case the radial metric function is an increasing function of the radial coordinate $r$, and its rate of increase is strongly dependent on the variations in the numerical values of $\psi _0$. Similarly to the previous cases, the string tension is a monotonically decreasing function of $r$, and it vanishes at a finite value of $r$, $r=R_s$, which sets the string radius. The string radius is weakly dependent on the variation of $\psi _0$, however, significant variations in $\sigma$ do appear for small values of $r$.

\begin{figure*}[htbp]
\centering
\includegraphics[width=8.3cm]{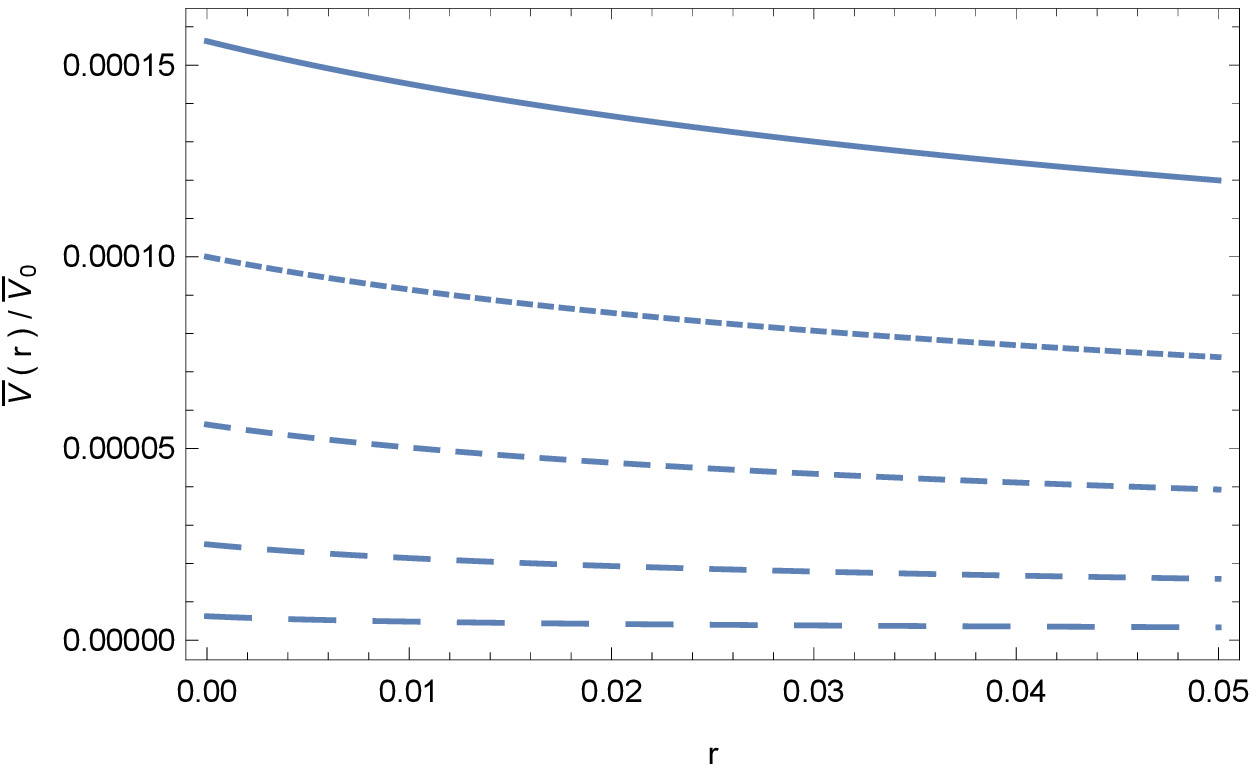}
\includegraphics[width=8.3cm]{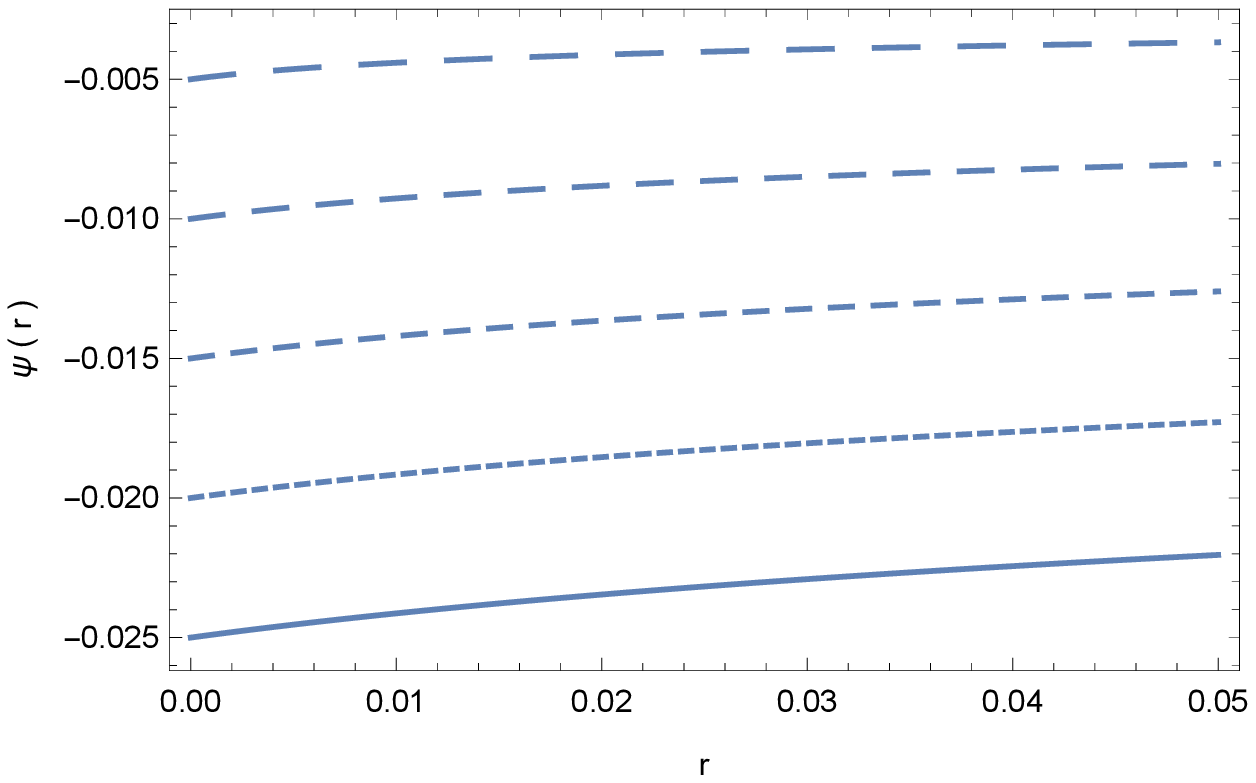}
\caption{Variations of the potential  $\bar{V}\left(\xi,\psi\right)=\bar{V}_0\xi ^2\psi ^2$ (left panel), and of the function $\psi $ (right panel) as a function of $r$ (with all quantities in arbitrary units) for the $\bar{V}(\xi)=\bar{V}_0\xi ^2\psi^2$ potential, for $\psi _0=-0.025$ (solid curve), $\psi _0=-0.020$ (dotted curve),
 $\psi _0=-0.015$ (short dashed curve), $\psi _0=-0.01$ (dashed curve), and $\psi _0=-0.005$ (long dashed curve), respectively. For $\bar{V}_0$ we have adopted the value $\bar{V}_0=10$, while the boundary conditions used to numerically integrate the field equations are $u_0=-0.01$, $\alpha_0=0.025$, $W(0)=0.10$, and $v_0=0.10$, respectively. }
\label{fig7}
\end{figure*}

The variations of the potential and of the function $\psi$ are represented in Fig.~\ref{fig7}. $\bar{V}$ is a slowly decreasing positive function of $r$, strongly dependent on the initial condition for $\psi'$. The function $\psi$ takes negative values, and show a strong dependence on $\psi _0$.

\begin{figure}[htbp]
\centering
\includegraphics[width=8.3cm]{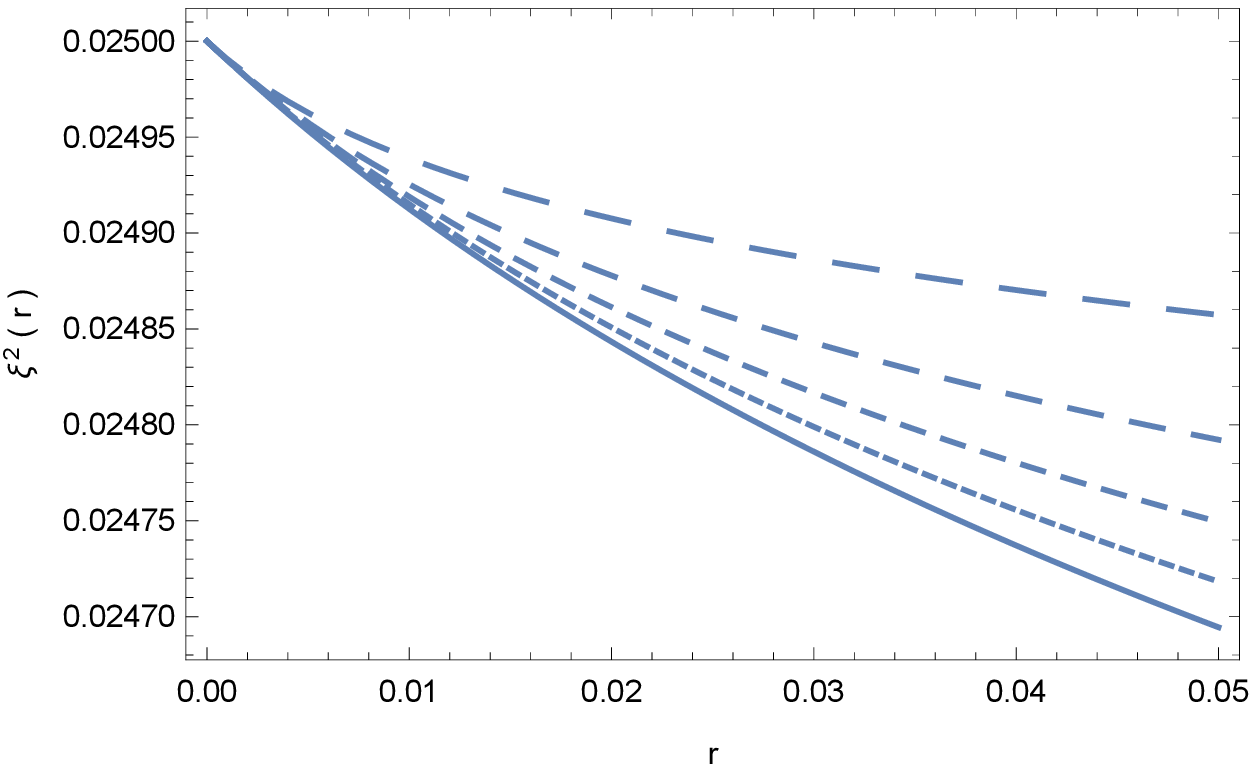}
\caption{Variation of $\xi^2$ as a function of $r$ (with all quantities in arbitrary units) for the $\bar{V}(\xi,\psi)=\bar{V}_0\xi^2 \psi^2$ potential, for $\psi _0=-0.025$ (solid curve), $\psi _0=-0.020$ (dotted curve),
 $\psi _0=-0.015$ (short dashed curve), $\psi _0=-0.01$ (dashed curve), and $\psi _0=-0.005$ (long dashed curve), respectively. For $\bar{V}_0$ we have adopted the value $\bar{V}_0=10$, while the boundary conditions used to numerically integrate the field equations are $u_0=-0.01$, $\alpha_0=0.025$, $W(0)=0.10$, and $v_0=0.10$, respectively. }
\label{fig8}
\end{figure}

The behavior of the function $\xi ^2(r)$ is depicted in Fig.~\ref{fig8}. $\xi ^2$ is positive for $r\in \left[0,R_s\right]$, and thus the physical nature of the gravitational coupling in the present model is guaranteed.   $\xi ^2$ is a monotonically decreasing function of $r$, and  its variation depends significantly on the numerical values of the initial conditions for $\psi '(0)$.

\section{Conclusions}

In this work we studied the existence and physical properties of local $U(1)$ cosmic strings in the context of the generalized hybrid metric-Palatini gravity. The theory is an extension to General Relativity, combining both metric and Palatini formalism. A main success of the theory is the possibility to generate long-range forces that pass the classical local tests of gravity at the Solar System level, thus avoiding some problematic features of the standard $f(R)$ theories. Another interesting advantage of the theory is that it admits an equivalent scalar-tensor representation, simplifying greatly the dynamical equations. 
The type of strings studied in this work are local gauge strings, using an approximation to the Vilenkin-prescribed energy-momentum tensor and different potential configurations.

The field equations determine the string tension $\sigma$. In order to solve these equations, one must choose the form of the potential and impose some appropriate boundary conditions as the core of the string, on the two scalar fields $\left(\xi ^2,\psi\right)$, on their derivatives, and for $W^2(0)$. Many types of cosmic string structures can be obtained by adopting some specific forms of $V$, and different sets of initial conditions, since the boundary conditions are, in this work, arbitrary and the system of equations is very sensitive to variations on the boundary conditions. 

In the case of the constant, nonzero, potential, the field equations can be solved exactly, and some simple expressions for the geometrical and physical parameters can be obtained. In this case, the string tension can be made positive by an appropriate choice of the potential. A solution with a constant string tension $\kappa ^2\sigma =\Lambda/2$ can also be constructed, as well as a solution having $W(r)=W_0r$, which can describe the standard general relativistic string if $W_0^2=1-8\pi G\mu$.
The string radius $R_s$ can be uniquely defined, and it is given in terms of the constant potential, as well as two integration constants. If we set the string tension to be positive defined in $r=0$, the string radius is also positive.

The large number of parameters of the models allows the construction of a large number of different numerical cosmic string models. However, we have restricted the set of parameters, as well as the physical nature of the solutions, by imposing three physical constraints, namely, that the string tension is positive inside the string, and it vanishes at the vacuum boundary, that the string must have a well defined and unique radius $R_s$, obtained from the condition $\sigma \left(R_s\right)=0$, and that $\xi ^2>0$, $\forall r\in \left[0,R_s\right]$. Even after these restrictions, a large variety of string models in generalized HMPG theory can be obtained.

In conclusion, in the present work we have considered specific cosmic string models that are solutions of the field equations of the generalized HMPG theory. Modified gravity theories may have profound implications on the formation, properties and structure of cosmic strings, interesting and important topological objects that may have been generated in the early Universe.  Hence, the theoretical investigations of strings in modified gravity models may therefore be a worthwhile pathway for future research.

\section{Acknowledgments}
FSNL acknowledges support from the Funda\c{c}\~{a}o para a Ci\^{e}ncia e a Tecnologia (FCT) Scientific Employment Stimulus contract with reference CEECINST/00032/2018, and the research grants No. UID/FIS/04434/2020, No. PTDC/FIS-OUT/29048/2017 and No.
CERN/FIS-PAR/0037/2019.  JLR was supported by the European Regional Development Fund and the programme Mobilitas Pluss (MOBJD647)


\section{References}

\begin{thebibliography}{99}
%\cite{Sotiriou:2008rp}
\bibitem{Sotiriou:2008rp}
T.~P.~Sotiriou and V.~Faraoni,
%``f(R) Theories Of Gravity,''
Rev. Mod. Phys. \textbf{82} (2010), 451-497
doi:10.1103/RevModPhys.82.451
[arXiv:0805.1726 [gr-qc]].
%2986 citations counted in INSPIRE as of 29 Oct 2021

%\cite{Khoury:2003aq}
\bibitem{Khoury:2003aq}
J.~Khoury and A.~Weltman,
%``Chameleon fields: Awaiting surprises for tests of gravity in space,''
Phys. Rev. Lett. \textbf{93} (2004), 171104
doi:10.1103/PhysRevLett.93.171104
[arXiv:astro-ph/0309300 [astro-ph]].
%1251 citations counted in INSPIRE as of 28 Oct 2021

%\cite{Khoury:2003rn}
\bibitem{Khoury:2003rn}
J.~Khoury and A.~Weltman,
%``Chameleon cosmology,''
Phys. Rev. D \textbf{69} (2004), 044026
doi:10.1103/PhysRevD.69.044026
[arXiv:astro-ph/0309411 [astro-ph]].
%1337 citations counted in INSPIRE as of 29 Oct 2021

%\cite{Olmo:2011uz}
\bibitem{Olmo:2011uz}
G.~J.~Olmo,
%``Palatini Approach to Modified Gravity: f(R) Theories and Beyond,''
Int. J. Mod. Phys. D \textbf{20} (2011), 413-462
doi:10.1142/S0218271811018925
[arXiv:1101.3864 [gr-qc]].
%462 citations counted in INSPIRE as of 29 Oct 2021

%\cite{Gomez:2020rnq}
\bibitem{Gomez:2020rnq}
D.~S\'aez-Chill\'on G\'omez,
%``Variational principle and boundary terms in gravity $Ã  la$ Palatini,''
Phys. Lett. B \textbf{814} (2021), 136103
doi:10.1016/j.physletb.2021.136103
[arXiv:2011.11568 [gr-qc]].
%8 citations counted in INSPIRE as of 11 Oct 2021

%\cite{Harko:2011nh}
\bibitem{Harko:2011nh}
T.~Harko, T.~S.~Koivisto, F.~S.~N.~Lobo and G.~J.~Olmo,
%``Metric-Palatini gravity unifying local constraints and late-time cosmic acceleration,''
Phys. Rev. D \textbf{85} (2012), 084016
doi:10.1103/PhysRevD.85.084016
[arXiv:1110.1049 [gr-qc]].
%138 citations counted in INSPIRE as of 18 Oct 2021

%\cite{Harko:2018ayt}
\bibitem{Harko:2018ayt}
T.~Harko and F.~S.~N.~Lobo,
%``Extensions of f(R) Gravity: Curvature-Matter Couplings and Hybrid Metric-Palatini Theory,''
%5 citations counted in INSPIRE as of 11 Oct 2021

%\cite{Capozziello:2015lza}
\bibitem{Capozziello:2015lza}
S.~Capozziello, T.~Harko, T.~S.~Koivisto, F.~S.~N.~Lobo and G.~J.~Olmo,
%``Hybrid metric-Palatini gravity,''
Universe \textbf{1} (2015) no.2, 199-238
doi:10.3390/universe1020199
[arXiv:1508.04641 [gr-qc]].
%94 citations counted in INSPIRE as of 18 Oct 2021

%\cite{Harko:2020ibn}
\bibitem{Harko:2020ibn}
T.~Harko and F.~S.~N.~Lobo,
%``Beyond Einstein\textquoteright{}s General Relativity: Hybrid metric-Palatini gravity and curvature-matter couplings,''
Int. J. Mod. Phys. D \textbf{29} (2020) no.13, 2030008
doi:10.1142/S0218271820300086
[arXiv:2007.15345 [gr-qc]].
%20 citations counted in INSPIRE as of 11 Oct 2021

%\cite{Weinberg:1967tq}
\bibitem{Weinberg:1967tq}
S.~Weinberg,
%``A Model of Leptons,''
Phys. Rev. Lett. \textbf{19} (1967), 1264-1266
doi:10.1103/PhysRevLett.19.1264
%13311 citations counted in INSPIRE as of 29 Oct 2021

%\cite{Salam:1959zz}
\bibitem{Salam:1959zz}
A.~Salam and J.~C.~Ward,
%``Weak and electromagnetic interactions,''
Nuovo Cim. \textbf{11} (1959), 568-577
doi:10.1007/BF02726525
%254 citations counted in INSPIRE as of 11 Oct 2021

%\cite{Salam:1964ry}
\bibitem{Salam:1964ry}
A.~Salam and J.~C.~Ward,
%``Electromagnetic and weak interactions,''
Phys. Lett. \textbf{13} (1964), 168-171
doi:10.1016/0031-9163(64)90711-5
%1504 citations counted in INSPIRE as of 22 Oct 2021

%\cite{Amaldi:1991zx}
\bibitem{Amaldi:1991zx}
U.~Amaldi, W.~de Boer, P.~H.~Frampton, H.~Furstenau and J.~T.~Liu,
%``Consistency checks of grand unified theories,''
Phys. Lett. B \textbf{281} (1992), 374-382
doi:10.1016/0370-2693(92)91158-6
%181 citations counted in INSPIRE as of 11 Oct 2021

%\cite{Jeannerot:2003qv}
\bibitem{Jeannerot:2003qv}
R.~Jeannerot, J.~Rocher and M.~Sakellariadou,
%``How generic is cosmic string formation in SUSY GUTs,''
Phys. Rev. D \textbf{68} (2003), 103514
doi:10.1103/PhysRevD.68.103514
[arXiv:hep-ph/0308134 [hep-ph]].
%331 citations counted in INSPIRE as of 11 Oct 2021

%\cite{Mermin:1979zz}
\bibitem{Mermin:1979zz}
N.~D.~Mermin,
%``The topological theory of defects in ordered media,''
Rev. Mod. Phys. \textbf{51} (1979), 591-648
doi:10.1103/RevModPhys.51.591
%294 citations counted in INSPIRE as of 11 Oct 2021

%\cite{Chuang:1991zz}
\bibitem{Chuang:1991zz}
I.~Chuang, B.~Yurke, R.~Durrer and N.~Turok,
%``Cosmology in the Laboratory: Defect Dynamics in Liquid Crystals,''
Science \textbf{251} (1991), 1336-1342
doi:10.1126/science.251.4999.1336
%200 citations counted in INSPIRE as of 19 Oct 2021

%\cite{Salomaa:1987zz}
\bibitem{Salomaa:1987zz}
M.~M.~Salomaa and G.~E.~Volovik,
%``Quantized vortices in superfluid He-3,''
Rev. Mod. Phys. \textbf{59} (1987), 533-613
[erratum: Rev. Mod. Phys. \textbf{60} (1988), 573-573]
doi:10.1103/RevModPhys.59.533
%134 citations counted in INSPIRE as of 11 Oct 2021

%\cite{Abrikosov:1956sx}
\bibitem{Abrikosov:1956sx}
A.~A.~Abrikosov,
%``On the Magnetic properties of superconductors of the second group,''
Sov. Phys. JETP \textbf{5} (1957), 1174-1182
%963 citations counted in INSPIRE as of 19 Oct 2021

%\cite{Higgs:1964pj}
\bibitem{Higgs:1964pj}
P.~W.~Higgs,
%``Broken Symmetries and the Masses of Gauge Bosons,''
Phys. Rev. Lett. \textbf{13} (1964), 508-509
doi:10.1103/PhysRevLett.13.508
%6653 citations counted in INSPIRE as of 29 Oct 2021

%\cite{Vilenkin:1981zs}
\bibitem{Vilenkin:1981zs}
A.~Vilenkin,
%``Gravitational Field of Vacuum Domain Walls and Strings,''
Phys. Rev. D \textbf{23} (1981), 852-857
doi:10.1103/PhysRevD.23.852
%997 citations counted in INSPIRE as of 25 Oct 2021

%\cite{Rosa:2018jwp}
\bibitem{Rosa:2018jwp}
J.~L.~Rosa, J.~P.~S.~Lemos and F.~S.~N.~Lobo,
%``Wormholes in generalized hybrid metric-Palatini gravity obeying the matter null energy condition everywhere,''
Phys. Rev. D \textbf{98} (2018) no.6, 064054
doi:10.1103/PhysRevD.98.064054
[arXiv:1808.08975 [gr-qc]].
%41 citations counted in INSPIRE as of 22 Oct 2021
\end{thebibliography}
\end{document}